\documentclass[12pt]{article}
\pdfoutput=1

\usepackage{epsfig}
\usepackage{amsfonts}
\usepackage{amssymb}
\usepackage{xcolor}

\begin{document}
\newcommand{\nc}{\newcommand}
\nc{\beq}{\begin{equation}} \nc{\eeq}{\end{equation}}
\nc{\beqa}{\begin{eqnarray}} \nc{\eeqa}{\end{eqnarray}}
\nc{\R}{{\cal R}}
\nc{\A}{{\cal A}}
\nc{\K}{{\cal K}}
\nc{\B}{{\cal B}}
\begin{center}

{\bf   HOW ONE CAN OBTAIN UNAMBIGUOUS PREDICTIONS FOR THE\\[0.3cm] S-MATRIX IN NON-RENORMALIZABLE THEORIES} \vspace{1.0cm}

{\bf \large D. I. Kazakov$^{1,2}$} \vspace{0.5cm}

{\it
$^1$Bogoliubov Laboratory of Theoretical Physics, Joint
Institute for Nuclear Research, Dubna, Russia.\\
$^2$Moscow Institute of Physics and Technology, Dolgoprudny, Russia\\
e-mail: kazakovd@theor.jinr.ru}
\vspace{0.5cm}

\abstract{The usual Bogolyubov $\R$-operation works in non-renormalizable theories in the same way as in renormalizable ones. However, in the non-renormalizable case, the counter-terms eliminating ultraviolet divergences do not repeat the structure of the original Lagrangian but contain new terms with a higher degree of fields and derivatives increasing from order to order of PT. If one does not aim to obtain finite off-shell Green functions but limits  oneself only to the finiteness of the S-matrix, then one can use the equations of motion and drastically reduce the number of independent counter-terms. For example, it is possible to reduce all counter-terms to a form containing only operators with four fields and an arbitrary number of derivatives. And although there will still be infinitely many such counter-terms, in order to fix the arbitrariness of the subtraction procedure, one can normalize the on-shell 4-point amplitude, which must be known for arbitrary kinematics, plus the 6-point amplitude at one point. All other multiparticle amplitudes will be calculated unambiguously. Within the framework of perturbation theory, the number of independent counter-terms in a given order is limited, so does the number of normalization conditions. The constructed counter-terms are not absorbed into the normalization of a single coupling constant, the Lagrangian contains an infinite number of terms, but after fixing the arbitrariness, it allows one to obtain unambiguous predictions for observables.}


\end{center}
\newpage
\section{Introduction}

The Standard Model in particle physics is based on renormalizable interactions. The idea of renormalizability served as a guiding line for the construction of the Standard Model, including the mechanism of spontaneous symmetry breaking.
At the same time, the theory of gravity belongs to a non-renormalizable class of theories and requires an adequate description. The reason why non-normalizable interactions are considered unacceptable is twofold. Firstly, ultraviolet divergences generate an infinite class of new structures, the fixation of which leads to infinite arbitrariness. Secondly, scattering amplitudes in such theories grow in a power-law manner, violating unitarity.

In this note, we touch on the first of these problems and propose how to obtain unambiguous predictions in non-normalizable theories. At the same time, we do not assert anything radically new, infinite arbitrariness remains infinite arbitrariness, but we propose how it can be fixed and get unambiguous predictions for the observables. The key objects in our consideration are the scattering amplitudes on the mass shell. Our task is to get finite and unambiguous answers for them. At the same time, we deliberately ignore Green's functions outside the mass shell, they can be infinite or ambiguous, which, however, does not affect the predictions for the observables.

For the sake of clarity of our consideration, we will constantly draw parallels with the renormalized case and emphasize where the difference appears. Our argument is based on the standard BPHZ $\R$-operation~\cite{BPHZ} aimed at eliminating UV divergences. It remains unchanged and divergences are still eliminated by the introduction of local counter-terms.
A new key point is the use of the equations of motion, which makes it possible to reduce the number of independent operators and limit it to a set of operators with a fixed number of external lines. In the renormalizable case, this leads to the usual counter-terms corresponding to the rerormalization of charges and masses, and in the non-renormalizable case, it allows one to fix infinite arbitrariness by setting \underline{one} scattering amplitude as a function of kinematic variables in a certain interval, and then calculate all other amplitudes in an unambiguous way. For comparison, in the renormalizable case, you need to set \underline{one} amplitude at \underline{one} kinematic point, which allows you to further calculate this and other amplitudes in arbitrary kinematics.

\section{Effective action and background field method}

For further discussion, it is convenient to use the background field method and calculate the divergent part of the effective action of $\Gamma(\Phi)$ defined through the functional integral as follows
~\cite{Bukh}:
\beq
e^{i\Gamma(\Phi)}=e^{iS(\Phi)}\int {\cal D}\phi e^{i\{S(\Phi+\phi)-S(\Phi)-\frac{\delta \Gamma(\Phi)}{\delta \Phi}\phi\}},
\label{eff}
\eeq
where integration is performed over quantum fields $\phi$, and the effective action depends on the background classical field $\Phi$. The classical action $S(\Phi)$ is the standard integral of the Lagrange function
\beq
S(\Phi)=\int d^Dx {\cal L}(\Phi).
\eeq
Expanding the classical action of $S(\Phi+\phi)$ in eq.(\ref{eff}) in a series over the quantum field $\phi$, we thereby generate vertices of interaction of the classical field $\Phi$ with the quantum field and the self-interaction of the quantum field. Next, we should consider connected Feynman diagrams, where the external fields are the background fields $\Phi$~\cite{Bukh}.

In the process of calculating the diagrams, ultraviolet divergences arise which are eliminated by introducing  local counter-terms that depend on the background field $\Phi$. The one-loop counter-term $\Delta S_1(\Phi)$ should be added to eq.(\ref{eff}) to eliminate the one-loop divergences. Further, when calculating two-loop divergences, it is necessary to shift the argument $\Phi\to \Phi+\phi$ in the one-loop counter-term $\Delta S_1(\Phi)$, as in the original action $S(\Phi)$. This will remove one-loop divergences in one-loop subgraphs that have both background and quantum external lines. After calculating the two-loop diagrams with background external lines, we find the two-loop divergences and obtain the two-loop local counter-term $\Delta S_2(\Phi)$, which also depends on the background field. This procedure is repeated in all the following orders of perturbation theory. As a result, at each stage we obtain a set of counter-terms that eliminate UV divergences in diagrams that depend on the external background field.

The effective action $\Gamma(\Phi)$ contains all the information about quantum corrections. In particular, to calculate the matrix elements of the S-matrix, it is enough to know the effective action depending on the fields $\Phi_{cl}$ satisfying the classical equations of motion~\cite{Bukh}
\beq
\frac{\delta {\cal L}(\Phi_{cl})}{\delta \Phi_{cl}}=0.
\eeq
Indeed, the generating functional for the elements of the S-matrix is the effective action $\Gamma(\phi_{in})$ containing all one-particle irreducible diagrams with external legs corresponding to free fields $\phi_{in}$
\beq
\phi_{in}(x)=\int\frac{d^3p}{\sqrt{(2\pi)^3\epsilon(p)}}\{a(\vec p)e^{-ipx}+a^*(\vec p)e^{ipx}\},
\eeq
here $p^2=0$ and $a(\vec p)$ and  $a^*(\vec p)$  are the free field classical amplitudes.
Then
\beq
S[a,a^*]=\Gamma[\phi_{in}]
\eeq
and differentiating the effective action over the classical amplitudes $a(\vec p)$ and $a^*(\vec p)$ followed by their zeroing gives the elements of the S-matrix.

Let us now consider $\Gamma[\Phi_{cl}]$, where $\Phi_{cl}$ is the solution of the complete classical equations
of motion. If we solve such an equation by iterations with respect
to the free equation of motion, then \cite{BD, ASF}
\beq
\Phi_{cl}(x)= \phi_{in}(x)+\int dy \Delta_{ret}(x-y)j(y), \label{sol}
\eeq
where $\Delta_{ret}(x-y)$ is the retarded Green function of the free field, and $j(x)=V'(x)$ is the nonlinear term in the complete classical equation of motion.
Then
\beq
\Gamma[\Phi_{cl}]=\Gamma[\phi_{in}]+\Delta\Gamma[\phi_{in}]=S[a,a^*]+\Delta S[a,a^*],
\eeq
and
$\Delta S[a,a^*]$ after differentiation by classical amplitudes with their subsequent zeroing does not contribute to the S-matrix, since as $t\to\infty$ in the right part of eq.(\ref{sol}) only the first term survives.
Then in fact $S[a,a^*]=\Gamma[\Phi_{cl}]$.

The last remark is the key for further consideration because it allows us to use the classical equations of motion to reduce the set of independent operators in the counter-terms, in both renormalizable and non-renormalizable cases.

Another argument in favour of using the equations of motion when calculating the elements of the S-matrix is the following~\cite{Hooft}.  Perform the field transformation
\beq \Phi(x)\to \Phi(x)+\Delta \Phi(x), \label{shift} \eeq
where a small variation of the field $\Delta \Phi(x)$  is proportional to the coupling  $\lambda$. Then the variation of the Lagrangian takes the form
\beq
{\cal L}[\Phi]\to{\cal L}[\Phi]+{\cal L}'[\Phi]\Delta\Phi+{\cal O}(\lambda^2).
\eeq

It is essential that the S-matrix elements with the proper external line renormalization factors are not influenced by the replacement of the fields (\ref{shift}).
From this it follows that any change of  $\Delta{\cal L}$, which is proportional to   ${\cal L}'[\Phi]$,   does not influence  the  S-matrix. In other words, one can use the equations of motion 
\beq
{\cal L}'[\Phi_{cl}]=0
\eeq
to simplify the expression for $\Delta{\cal L}$.

\section{Reduction of operators on equations of motion}

Consider a set of local counter-terms that eliminate all ultraviolet divergences in an effective action in a fixed order of perturbation theory for an arbitrary background field. Symbolically, they can be written as
\beq
\Delta {\cal L}= \sum_{i=1}^N z_i O_i(\Phi), \label{counterterm}
\eeq
where $z_i$ are some regularized constants, and $O_i$ are operators constructed from fields and their derivatives, not necessarily repeating the original Lagrangian.

Let us now take the equations of motion for all fields $R_j=0$, which relate various operators to each other.
For example, in theory with Lagrangian
\beq
{\cal L}=\frac 12 (\partial_\mu \phi)^2-\frac 12 m^2\phi^2-\frac{\lambda}{4!}\phi^4 \label{phi4}
\eeq
the equation of motion has the form $R_1=\partial^2 \phi +m^2 \phi+\lambda/3! \phi^3=0$ and connects three operators $\phi\partial^2 \phi$, $m^2 \phi^2$ and $\frac{\lambda}{3!}\phi^4$.

Then eq.(\ref{counterterm}) can be rewritten as
\beq
\Delta {\cal L}=\sum_{i =1}^{N}\sum_{j=1}^{M=N-K} c_{ij}z_iO_j(\Phi)+\sum_{j=1}^K z_j\Phi_jR_j (\Phi)\label{counterterm2}
\eeq
where a subclass of $M$ independent operators not reducible to each other on the equations of motion is allocated.
We emphasize that these counter-terms eliminate all UV divergences. Now, if you go to the mass shell, eq.(\ref{counterterm2}) will take the form
\beq
\Delta \tilde{\cal L}= \sum_{i =1}^{N}\sum_{j=1}^{M=N-K} c_{ij}z_i \tilde O_j(\Phi_{cl}), \label{counterterm3}
\eeq
where the operators $\tilde O_i$ are constructed from fields satisfying the equations of motion. These counter-terms ensure finiteness of amplitudes only on the mass shell.

In the above example (\ref{phi4}) of the $\phi^4$  theory in four dimensions we have, respectively,
\beqa
\Delta {\cal L}&=& -z_1 \frac 12 \Phi \partial^2 \Phi -\frac 12 z_2 m^2 \Phi^2-z_4 \frac{\lambda}{4!}\Phi^4, \\ 
\Delta {\cal L}&=&-\frac 12 (z_2-z_1) m^2 \Phi^2 -(z_4-2 z_1) \frac{\lambda}{4!} \Phi^4-\frac{z_1}{2}\Phi(\partial^2\Phi+m^2\Phi+\frac{\lambda}{6}\Phi^3),\\
\Delta \tilde{\cal L}&=&-\frac 12 (z_2-z_1) m^2 \Phi^2 -(z_4-2z_1)  \frac{\lambda}{4!} \Phi^4,\label{red}
\eeqa
where two operators without derivatives are chosen as independent operators. Note that the resulting combinations of counter-terms in eq.(\ref{red}) exactly correspond to the renormalization of the Lagrangian parameters, namely $m^2$ and $\lambda$, taking into account the renormalization of the fields: $z_m=z_2-z_1,
z_\lambda=z_4-2 z_1$.

As another example, consider the well-known quantum electrodynamics. The Lagrangian has the form
\beq
{\cal L}=-\frac 14 F_{\mu\nu}F_{\mu\nu}+\bar\psi (i\partial_\mu\gamma^\mu-m+e A_\mu\gamma^\mu)\psi,
\eeq
from where we get the equations of motion
\beqa
\partial_\mu F_{\mu\nu}+e\bar\psi \gamma_\mu\psi=0, \\
(i\partial_\mu \gamma^\mu-m+e A_\mu\gamma^\mu)\psi=0.
\eeqa
Since quantum electrodynamics refers to renormalizable theories, the counter-terms repeat the structure of the original Lagrangian and have the form
\beq
\Delta{\cal L}=-z_3\frac 14 F_{\mu\nu}F_{\mu\nu}+z_2\bar\psi i\partial_\mu \gamma^\mu\psi-z'm\bar\psi\psi+
z_1e \bar\psi A_\mu\gamma^\mu\psi,
\eeq
which can be rewritten as
\beqa
\Delta{\cal L}=-(z'-z_2)m\bar\psi\psi+
(z_1-z_2-\frac 12 z_3)e \bar\psi A_\mu\gamma^\mu\psi\\
-z_3\frac 14 (F_{\mu\nu}F_{\mu\nu}+2e\bar\psi \gamma_\mu\psi)+z_2\bar\psi( i\partial_\mu \gamma^\mu -m+e A_\mu\gamma^\mu)\psi,
\eeqa
and which, taking into account the equations of motion, gives
\beq
\Delta\tilde{\cal L}=-(z'-z_2)m\bar\psi\psi+
(z_1-z_2-\frac 12 z_3)e \bar\psi A_\mu\gamma^\mu\psi.
\eeq
As can be seen, in the case of QED, combinations of counter-terms also arise that exactly correspond to the renormalization of the Lagrangian parameters, namely $z_m=z'-z_2,
z_e=z_1- z_2-z_3/2$.

Thus, taking into account the equations of motion, there is a reduced, albeit ambiguous, irreducible set of operators (two instead of three in theory $\phi^4_4$ and two instead of four in QED) and, accordingly, the arbitrariness that must be fixed when eliminating UV divergences decreases. 

\section{From renormalizable to non-renormalizable theories: the structure of counter-terms}

The examples considered so far relate to renormalizable theories. In this case, the counter-terms reproduce the structure of the original Lagrangian; there is a finite number of such structures, and taking into account the equations of motion reduces their number, so that there are exactly as many structures as there are independent parameters (masses and coupling constants) in the Lagrangian. In the non-renormalizable case, the number of such structures increases from order to order of perturbation theory; they do not repeat the original Lagrangian and as a result there is an infinite number of counter-terms. Taking into account the equations of motion in this case also makes it possible to reduce their number, but there is still an infinite number of counter-terms of a certain type. Let us demonstrate this by using the massless $\phi^4$ theory in $6$ dimensions as an example.

There are non-zero 4, 6, 8, etc. point amplitudes. Let us start with a 4-point amplitude. It depends on the Mandelstam variables $s,t,u$ and we put all external momenta on the mass shell, i.e. $p_i^2=0$. Further on, we use dimensional regularization to work with UV divergences. Then in one loop we have the following expression
for the UV counter-term in the momentum representation
\beq
\Delta{\cal L}_1 \sim \lambda^2(s+t+u)\Phi^4(\frac 1\epsilon +c_ {11}),
\eeq
where $c_{11}$ is an arbitrary subtraction constant. In the coordinate representation, this corresponds to the counter-term
\beq
\Delta{\cal L}_1 \sim \lambda^2\partial^2 \Phi^2 \Phi^2(\frac 1\epsilon +c_{11}),
\eeq

If we continue to move along loops and still consider the 4-point amplitude, then local counter-terms of an increasingly complex structure containing a larger number of derivatives will arise. In addition, starting with one loop, divergences will appear in 6-point amplitudes; starting with two loops, in 8-point amplitudes, etc. New counter-terms will contain an increasing number of external fields along with derivatives. As discussed above, the constructed counter-terms depend on the external background field and in each order of perturbation theory are local functions, i.e. all fields are taken at one space-time point and contain a finite number of derivatives. The number of fields and derivatives is determined by dimensional counting: in six dimensions the total mass dimension of the counter-terms is equal to 6, the dimension of the field is 2 and that of the coupling  is minus 2.

Schematically, in the coordinate representation, we have the following loop expansion:
\begin{eqnarray*}
\hspace*{-1cm}\Delta {\cal L}&=&
\lambda^2\partial^2\Phi^2 \Phi^2 + \lambda^3[\partial^4\Phi^2 \Phi^2 \ \ \ \ + \ \ \ \ \ \ \ \partial^2 \Phi^2 \partial^2\Phi^2 )] +\hspace*{0.8cm} \lambda^4[...]+\hspace*{0.8cm}\lambda^5[...]\\
&&(\frac 1\epsilon +c_{11}) \ \ \ \ \ \ \ (\frac{1}{\epsilon^2} +\frac 1\epsilon+c_{12}) \ \ \ (\frac{1}{\epsilon^2} +\frac 1\epsilon+c_{13}) \\&& \hspace*{1.8cm} \searrow \hspace*{2.8cm} \searrow \hspace*{5.8cm} \searrow \\
&&\hspace*{2.4cm} \lambda^3\Phi^6 \hspace*{0.5cm}\ \ \ + \hspace{0.5cm} \lambda^4[\partial^2\Phi^4 \Phi^2 \ \ \ \ \ +\ \ \ \ \ \partial^2 \Phi^2 \Phi^4]\\
&&\hspace*{2cm} (\frac 1\epsilon +c_{21}) \ \ \ \ \ \ \ \ \ \ \ \ (\frac{1}{\epsilon^2} +\frac 1\epsilon+c_{22}) \ \ \ (\frac{1}{\epsilon^2} +\frac 1\epsilon+c_{23}) \\ &&\hspace*{8.8cm} \searrow \hspace*{3.8cm} \searrow \\
&&\hspace*{9.2cm} \lambda^5\Phi^8\\
&&\hspace*{9.2cm} (\frac{1}{\epsilon^2} +\frac 1\epsilon+c_{32}),
\end{eqnarray*}
where the first row corresponds to a 4-point amplitude, the second to a 6-point amplitude, etc. The arrows point to the
operators connected by the equations of motion. If we now take into account the equations of motion, we get a reduced set of counter-terms containing only terms with 4 external lines
\begin{eqnarray*}
\hspace*{-1cm}\Delta \tilde{\cal L}&=&
\lambda^2\partial^2\Phi^2 \Phi^2 (\frac 1\epsilon +c_{11}+c_{21}) \\
&+& \lambda^3[\partial^4\Phi^2 \Phi^2 (\frac{1}{\epsilon^2} +\frac 1\epsilon+c_{12}+c_{22}+c_{32}) + \partial^2 \Phi^2 \partial^2\Phi^2 (\frac{1}{\epsilon^2} +\frac 1\epsilon+c_{13}+c_{23})] + ...
\end{eqnarray*}

As one can see, everything is reduced to operators containing only four powers of the field but increasing, from order to order, the number of derivatives.
Thus, a two-dimensional array of operators, when taking into account the equations of motion, is reduced to one-dimensional, so that all arbitrariness in the normalization of operators also becomes one-dimensional, although infinite. Therefore, to fix the arbitrariness of counter-terms on the mass shell, it is sufficient to set the normalization of the four-point amplitude, and the remaining multi-leg amplitudes will be calculated unambiguously. At the same time, the four-point amplitude must be known for arbitrary kinematics in a certain range of values of $s$ and $t$. It then continues analytically for all values. This means that the four-point amplitude is not predicted, but is used only for normalization. However, if we work within the framework of perturbation theory, then in a given order of PT the number of counter-terms is limited and a finite number of normalization conditions is required. This will allow one to fix all arbitrariness in a given order of PT and get unambiguous predictions for the remaining amplitudes.

\section{Illustration}

For comparison, let us start with the massless $\phi^4_4$ theory. This theory is renormalizable, the coupling constant is dimensionless, divergences are present only in the 4-point functions. The corresponding amplitude has the form:
\beq
A_4= \lambda+A  \lambda^2(\frac 3 \epsilon - \log(s/\mu)-\log(t/\mu)-\log(u/\mu)) + ...
\eeq
The counter-term in the coordinate representation repeats the structure of the original Lagrangian and contains one arbitrary subtraction constant
$$\Delta {\cal L} = A \frac{ \lambda^2}{4!}\Phi^4 (-\frac 3\epsilon-c),$$
so that the finite part takes the form
\beq
A_4^{finite}= \lambda+A \lambda^2(- \log(s/\mu)-\log(t/\mu)-\log(u/\mu)-c)+... \label{4}
\eeq
The constant $c$ can be included in the definition of the parameter $\mu$.

Now we fix the arbitrariness: we define the coupling constant $ \lambda$ from the normalization of the 4-point function as a solution of the equation
\beq
A_4^0= \lambda+A \lambda^2(\log(s_0/\mu)-\log(t_0/\mu)-\log(u_0/\mu)-c).
\eeq
Substituting a perturbative solution
\beq
 \lambda=A_4^0+(A_4^0)^2 A(\log(s_0/\mu)-\log(t_0/\mu)-\log(u_0/\mu)-c).
\eeq
in equation (\ref{4}), we get
\beq
A_4^{finite}=A_4^0+(A_4^0)^2 A(- \log(s/s_0)-\log(t/t_0)-\log(u/u_0))+...
\eeq
so all arbitrariness disappears and we have an unambiguous expression for the 4-point amplitude expressed in terms of its value at one point.

Let us now turn to the massless $\phi^4_6$ theory.  The peculiarity here is that the divergence of the symmetric four-point amplitude is proportional to $s+t+u=0$ and, accordingly, there is no one-loop divergence on the mass shell. Due to the structure of the $\R$-operation, in this case, all leading multi-loop divergences also vanish~\cite{we}. However, the subleading ones remain. This means that in two loops there is no senior pole $1/\epsilon^2$, and there is only a junior pole $1/\epsilon$, in three loops there is no pole $1/\epsilon^3$ but there are junior poles $1/\epsilon^2$ and $1/\epsilon$, etc.

Thus, in the massless $\phi^4_6$ theory  we have the following expression for the 4-point amplitude in the one-loop approximation
\beqa A_4&=&\lambda+A \lambda^2 (\frac{s+t+u}{\epsilon}- s\log(s/\mu)-t\log(t/\mu)-u\log(u/\mu))+...\\
&=&
\lambda+A\lambda^2(- s\log(s/u)-t\log(t/u))+...
\eeqa
However, at D=6, the 6-point amplitude also diverges, starting with one loop. Schematically,
\beq
A_6=B \lambda^3 (\frac 1\epsilon - \log Q/\mu)+...,
\eeq
where $Q$ is some generalized momentum from the 6-point function.
Thus, the counter-terms in the coordinate representation have the form
\beq
 \Delta {\cal L}= \partial^2\Phi^2 \Phi^2 A\lambda^2 \times 0+B\lambda^3\Phi^6(-\frac 1\epsilon -c).
 \eeq
Let us return now to the finite parts. We get, respectively,
\beqa 
A_4^{finite}&=& \lambda+A\lambda^2(- s\log(s/u)-t\log(t/u)), \\
A_6^{finite}&=&B \lambda^3 (- \log Q/\mu-c).
\eeqa
The constant $c$ can again be included in the definition of the parameter $\mu$.

Now we fix the arbitrariness:
we determine the coupling constant $\lambda$ from the normalization of the 4-point amplitude:
\beq
A_4^0=\lambda+A\lambda^2(-s_0\log(s_0/u_0)-t_0\log(t_0/u_0)), \label{6}
\eeq
and the scale $\mu$ from the  normalization of the 6-point amplitude:
\beq
A_6^0=B \lambda^3 ( - \log Q_0/\mu),
\eeq
where the constant $\lambda$ is taken from the solution of eq.(\ref{6}).
Futher on, in higher loops, we will write down all counter-terms as 4-point ones and fix the arbitrariness by imposing new conditions on the 4-point amplitude, while the condition on the 6-point amplitudes will remain the same, only the equation on $\mu$ will become more complicated. For the 8-point and other multi-leg amplitudes, there will be no arbitrariness anymore.

To summarize: to fix the complete arbitrariness in the amplitudes on the mass shell in the $\phi^4_6$ theory,
it is necessary to impose an infinite number of conditions on the 4-point amplitude (in the finite order of PT, the number of conditions imposed on the 4-point amplitude is also finite) and one on the 6-point amplitude. Thus, the 4-point amplitude is used only for normalization and is not predicted by theory; for the 6-point amplitude, it is necessary to fix only the value at one point and the remaining values are unambiguously predicted. All other amplitudes are uniquely calculated by perturbation theory.

Let us now consider the massless $\phi^4_8$ theory. The difference is that now in one loop we have a divergence proportional to $s^2+t^2+u^2\neq 0$ and it does not disappear on the mass shell,
and the 6-point and 8-point functions also diverge.
For the amplitudes we have the following expressions, respectively,
\beqa A_4&=&\lambda+A \lambda^2 (\frac{s^2+t^2+u^2}{\epsilon}- s^2\log(s/\mu)-t^2\log(t/\mu)-u^2\log(u/\mu))+...\\
A_6&=&B \lambda^3 Q^2(\frac 1\epsilon - \log Q/\mu)+...,\\
A_8&=& E\lambda^4 (\frac 1\epsilon -\log P/\mu)+...
\eeqa
where $P$ is some generalized momentum from the 8-point amplitude.
The counter-terms in the coordinate representation therefore have the form
\beq
 \Delta {\cal L}= \partial^4\Phi^2 \Phi^2 A\lambda^2(-\frac 1\epsilon -c_1)+ B\lambda^3 \partial^2\Phi^2 \Phi^4(-\frac 1\epsilon -c_2)+
E\lambda^4 \Phi^8(-\frac 1\epsilon -c_3).
\eeq
Taking into account the equations of motion, we choose them in the form
\beq 
\Delta\tilde{\cal L}=\partial^4\Phi^2 \Phi^2 A\lambda^2(-\frac{A+B+E}{\epsilon} -c_1-c_2-c_3)=\partial^4\Phi^2\Phi^2 A\lambda^2(-\frac{A+B+E}{\epsilon} -c),
\eeq
i.e. there is only one undefined subtraction constant $c$.
The finite parts then become, respectively,
\beqa A_4^{finite}&=&\lambda+A \lambda^2 (- s^2\log(s/\mu)-t^2\log(t/\mu)-u^2\log(u/\mu)-c(s^2+t^2+u^2))+...\\
A_6^{finite}&=&B \lambda^3 Q^2( - \log Q/\mu)+...,\\
A_8^{finite}&=& E\lambda^4 ( - \log P/\mu)+...
\eeqa
Now to fix the arbitrariness, we will impose two conditions on the 4-point amplitude and one condition on the 6-point amplitude.
The first fix $\lambda$ and $c$ ($\mu$ will fall out); and the second, $\mu$. Further on, as in the case of
$\phi^4_6$, it is necessary to impose new conditions on the 4-point amplitude, and the condition on the 6-point amplitude will remain the same.

To summarize: to fix the complete arbitrariness in the amplitudes on the mass shell in the $\phi^4_8$ theory,
it is necessary to impose an infinite number of conditions on the 4-point amplitude (in the finite order of PT, the number of conditions imposed on the 4-point amplitude is finite) and one on the 6-point amplitude, and all other amplitudes are uniquely calculated, i.e., the situation with divergences almost completely repeats the case of $\phi^4_6$.

\section{Counter-terms on the mass shell}

Note that if you aim at calculating the S-matrix, then you can stay on the mass shell, as was done when calculating scattering amplitudes in the method of unitary cuts, where all calculations are carried out on the mass shell and multi-loop diagrams are obtained by gluing diagrams with fewer loops with the ends on the mass shell~\cite{Reviews_Ampl_General}.
This property is essential when calculating diagrams, allowing one to immediately impose mass shell conditions on external momenta, which leads to simplification of calculations but is not principal for the procedure of eliminating divergences and fixing arbitrariness.

We will examplify how divergences can be eliminated in multi-loop diagrams on the mass shell using only the corresponding lower-order counter-terms calculated on-shell.

\begin{figure}[h]
\begin{center}
\includegraphics[scale=0.50]{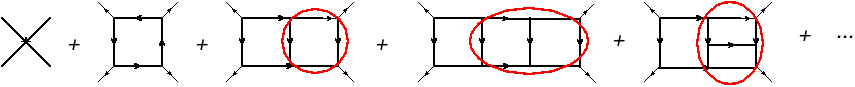}
\caption{Diagrams giving a contribution to the four-point amplitude. Divergent subgraphs are highlighted}
\label{amp}
\end{center}
\end{figure}

Consider the scattering amplitude $2\to 2$ on the  mass shell in a theory with triple vertices of the type that occurs in supersymmetric gauge theory~\cite{Boxes,we3}. We have the sequence of diagrams shown in Fig.\ref{amp}.

Counter-terms constructed on the basis of the $\R$-operation due to the Bogolyubov-Parasyuk theorem always have a local character~\cite{BP}. This means that the one-loop counter-term is a polynomial of degree equal to the UV divergence of the diagram of the kinematic variables. In this case, these are the Mandelstam variables $s,t,u$ and the squares of the external momenta $p_i^2$. In the massless theory, for momenta lying on the mass shell, all $p_i^2=0$, so the counter-terms on the mass shell do not contain such terms. At the same time, a two-loop diagram contains a one-loop subgraph whose two ends do not lie on the mass shell and one would expect that an off-shell counter-term would be needed to eliminate the UV divergence. However, this is not the case.

\begin{figure}[h]
\begin{center}
\includegraphics[scale=0.50]{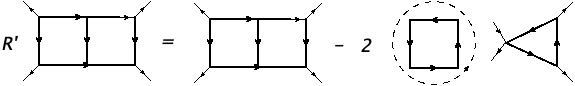}
\caption{Action $\R$-operation on a two-loop diagram}
\label{Rop}
\end{center}
\end{figure}

Indeed, the action of the $\R'$-operation on the two-loop diagram has the form shown in Fig.\ref{Rop},
where the one-loop subgraph surrounded by the dotted line corresponds to the counter-term, which, as already indicated, is a polynomial in $s,t,u$ and $p_i^2$. These momenta should be integrated along the remaining triangle. In this case, the terms of the polynomial proportional to $p_i^2$
cancel one of the lines in the diagram so that it takes the form
\begin{center}
\includegraphics[scale=0.45]{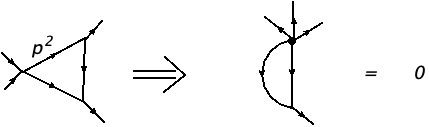}
\end{center}
and vanishes because it depends only on one argument equal to zero on the mass shell.

Exactly the same situation occurs in three-loop diagrams. Indeed, following the procedure of the
${\cal R}$-operation, the counter-terms corresponding to a given graph $G$ are given by:
\beq
\Delta {\cal L}(G)= -\K\R' G, \label{ZZ}
\eeq
whereas the incomplete ${\cal R}$-operation, called the ${\cal R}'$-operation, subtracts all subdivergences in each subgraph $G_\gamma$ of the graph $G$ (the operator $\K$ selects the singular part)~\cite{Rop} 
\begin{equation}
{\cal R}' G= G-\sum_\gamma \K{\cal R}'_\gamma  G_{/\gamma}+\sum_{\gamma,\gamma'}\K{\cal R}'_\gamma \K{\cal R}'_{\gamma'} G_{/\gamma\gamma'}- ...
\end{equation}
Here $\K{\cal R}'_\gamma$ stands for the counter term of the graph $\gamma$ and $G_{/\gamma}$ for the remaining part of the graph $G$ after shrinking the subgraph $\gamma$ to a point. The sum goes over all 1PI UV-divergent subgraphs of the given diagram and multiple sums include only non-intersecting subgraphs.

 This means that for the two-loop subgraphs highlighted in Fig.\ref{Rop} we substitute the corresponding local counter-terms for the entire circled subgraph, i.e. we use the expression obtained at the previous stage ${\cal KR}'_2$.

A similar argument can be extended to an n-loop diagram: if the (n-1)-loop subgraph 
 is located at the edge of the diagram, then the situation is not different from the case considered for n=2. If the (n-1)-loop counter-term contains a product of lower-order counter-terms, then the one-loop diagram may be in the middle. Then the term containing $p_i^2$ cancels the inner line that turns the diagram into a tadpole, which is also zero.
\begin{center}
\includegraphics[scale=0.50]{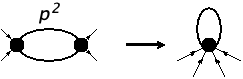}
\end{center}
Thus, those polynomial terms that are proportional to $p_i^2$ and do not disappear outside the mass shell  simply do not contribute to the calculation of amplitudes on the mass shell.
Hence, being limited to amplitudes on the mass shell, we can consider only counter-terms on the mass shell.

\section{Discussion}

Thus, the use of the equations of motion makes it possible to reduce the number of independent operators and limit it to a set of operators with a fixed number of external lines. In the non-renormalizable case, this makes it possible to fix infinite arbitrariness by setting \underline{one} 4-point scattering amplitude as a function of kinematic variables in a certain interval (plus one condition for the 6-point amplitude) and then calculate all other amplitudes in an unambiguous way. This opens up the possibility to work with non-renormalizable theories and obtain unambiguous predictions. In this case, the initial Lagrangian acquires an infinite number of new structures, which are \underline{unambiguously} fixed by the normalization conditions. There are a finite number of such structures in each  order of PT. The example we have considered with one field and one type of interaction is only an illustration of the general approach.

Thus, the procedure for obtaining predictions for observables - scattering amplitudes on the mass shell - is as follows: One has to calculate the amplitudes directly on the mass shall, i.e. with correct kinematics. After the calculation, it is necessary to subtract all divergences with
arbitrary subtraction constants and then fix these constants by imposing conditions on the scattering amplitudes. In any theory, it is possible to use the amplitude of $2\to 2$ scattering to fix the arbitrariness. Then all other amplitudes will be determined unambiguously.

Two observations should be made here. The first concerns arbitrariness in the choice of normalization conditions, which in renormalizable theories corresponds to the choice of a subtraction scheme. Unlike renormalizable theories, where when switching from scheme to scheme the renormalized coupling constants are multiplied by a constant multiplier $Z$, in non-renormalized theories, renormalization is not multiplicative, but depends on kinematics and has an integral character~\cite{Kaz}. Nevertheless, even in this case, the transition from one subtraction scheme to another is carried out by multiplying all counter-terms of a given order by a finite factor. The unambiguity of the answer, as in renormalizable theories, is achieved due to the fact that while expressing the multi-leg amplitude through a four-point one with fixed kinematics, the entire dependence on the arbitrariness of subtraction is cancelled.

The second remark concerns the growth of amplitudes with energy in each fixed order of perturbation theory. This is because the coupling constant has a negative dimension, which is compensated by degrees of energy. A typical term of a PT series looks like
\beq
g^n s^n \log s \label{PT}
\eeq
and it grows with energy. The way out of this situation may consist in summing up all the leading contributions of the type (\ref{PT}) in all orders of perturbation theory, which is achieved using generalized
renormalization group equations~\cite{we,PL2}. In this case, the question of the behavior of amplitudes with increasing energy should be raised after summing the leading asymptotics. Note at the same time that the approximation of the leading logarithms is universal and does not depend on the arbitrariness in the subtraction scheme in both renormalizable and non-renormalizable theories.

\section*{Acknowledgments}
This work is supported by the Russian Science Foundation grant  \# 21-12-00129. The author is grateful to I.L.Buchbinder for numerous useful discussions and clarifications regarding the formalism of effective action.

\end{document}